\documentclass[a4paper,12pt]{article}
\usepackage{amssymb}
\usepackage{amsfonts}
\usepackage{amsmath}
\usepackage[font={small},labelsep=period]{caption}
\usepackage[square,numbers,sort&compress,comma]{natbib}
\usepackage{hyperref}

\hypersetup{colorlinks=true, linkcolor=blue, citecolor=blue, urlcolor=blue}
\gdef\unit#1{\mathord{\thinspace\rm #1}}
\input{tcilatex}
\begin{document}

\author{M. S. Bieniek, D. F. N. Santos, P. G. C. Almeida, and M. S. Benilov$%
^{a}$\smallskip \\
%EndAName
Departamento de F\'{\i}sica, Faculdade de Ci\^{e}ncias Exatas e da
Engenharia,\\
Universidade da Madeira, Largo do Munic\'{\i}pio, 9000 Funchal, Portugal,
and\smallskip \\
Instituto de Plasmas e Fus\~{a}o Nuclear, Instituto Superior T\'{e}cnico,\\
Universidade de Lisboa, 1041 Lisboa, Portugal\smallskip \\
$^{a}$ email \textit{benilov@uma.pt}}
\title{Bifurcations in the theory of current transfer to cathodes of dc
discharges and observations of transitions between different modes}
\date{}
\maketitle

\begin{abstract}
General scenarios of transitions between different spot patterns on
electrodes of dc gas discharges and their relation to bifurcations of
steady-state solutions are analyzed. In the case of cathodes of arc
discharges, it is shown that any transition between different modes of
current transfer is related to a bifurcation of steady-state solutions. In
particular, transitions between diffuse and spot modes on axially symmetric
cathodes, frequently observed in the experiment, represent an indication of
the presence of pitchfork or fold bifurcations of steady-state solutions.
Experimental observations of transitions on cathodes of dc glow
microdischarges are analyzed and those potentially related to bifurcations
of steady-state solutions are identified. The relevant bifurcations are
investigated numerically and the computed patterns are found to conform to
those observed in the course of the corresponding transitions in the
experiment.
\end{abstract}

\section{Introduction}

Luminous spots on electrodes of direct current glow and arc discharges and
self-organized patterns of spots represent a very interesting phenomenon,
which is also important for applications. The presence, or not, of spots on
electrodes is a key point for the operation of any arc device.
Self-organized patterns appearing on cathodes of dc glow microdischarges are
sources of excimer radiation \cite{Takano2006,Zhu2007}. Self-organized
patterns on liquid anodes of atmospheric pressure glow microdischarges have
been shown to produce a nontrivial cancer-inhibiting effect \cite{Chen2017a}.

The theoretical description of spots and spot patterns on electrodes of dc
glow and arc discharges is based on the multiplicity of solutions: an
adequate theoretical model must in some cases allow multiple steady-state
solutions to exist for the same conditions (in particular, for the same
discharge current $I$), with different solutions describing the spotless
(diffuse) mode of current transfer and modes with different spot
configurations.

Some of the multiple solutions may merge, or become identical at certain
values of the control parameter; a bifurcation, or branching, of solutions.
Bifurcations of different kinds of steady-state solutions have been
encountered in the theory and modelling of current transfer to cathodes of
dc glow and high-pressure arc discharges \cite{2009f,2014b}. An
understanding of these bifurcations is crucial for the computation of the
whole pattern of multiple solutions and an analysis of their stability. The
existence of multiple solutions and their bifurcations, in the case of
current transfer to cathodes of DC glow discharges, is a consequence of a
strong positive feedback, which originates in the increasing dependence of
the rate of ionization on electric field. In the case of current transfer to
cathodes of arc discharges, the existence of bifurcations is a result of a
strong positive feedback originating in the dependence of the density of the
energy flux from the plasma to the cathode surface on the surface
temperature \cite{2014b}.

In contrast, multiple steady-state solutions describing different modes of
current transfer to anodes of glow microdischarges computed recently \cite%
{2017g} do not reveal bifurcations. The existence of multiple solutions in
this case is related to the change of sign of the anode sheath voltage.

Thus, the existence, or not, of bifurcations of steady-state solutions,
describing different modes of current transfer to electrodes of dc glow and
arc discharges, is related to the underlying physics and is therefore of
significant interest. Unfortunately, the question of whether bifurcations
exist has not been addressed in experimental publications. (Although there
are interesting results concerning bifurcations in the pattern of
oscillations developing in a dc-driven semiconductor-gas discharge system 
\cite{Mansuroglu2017}; see also \cite{Raizer2013,Rafatov2016} and review 
\cite{Purwins2014}.) It is therefore of interest to analyze available
experimental observations of different modes of current transfer to
electrodes of dc glow and arc discharges with the aim to eventually identify
bifurcations.

The outline of the paper is as follows. In Sec. \ref{Transitions between
different spot patterns}, the general scenarios of changes between modes on
electrodes of dc gas discharges and their relation to bifurcations of
steady-state solutions are analyzed. Transitions of modes on cathodes of
arc, and dc glow, discharges are considered in Secs.\ \ref{Cathodes of arc
discharges} and \ref{Bifurcations on cathodes of dc glow discharges},
respectively. The conclusions are summarized and directions of future work
are discussed in Sec.\ \ref{Conclusion}.

\section{Scenarios of transitions between different modes of current
transfer to electrodes of dc discharges and their relation to bifurcations}

\label{Transitions between different spot patterns}

Bifurcations of steady-state solutions manifest in experiments as
transitions between modes with different spot patterns, which occur as the
discharge current $I$ is varied.

One can distinguish two scenarios for transitions between modes with
different spot patterns. First, there are quasi-stationary, i.e.,
continuous, and, consequently, reversible transitions between states where
distributions of luminosity over the electrode surface possess different
symmetries. Second, there are transitions that occur abruptly even for very
small variations of $I$. Let us consider first the quasi-stationary
transitions. All parameters of the discharge, including the discharge
voltage $U$, vary with $I$ continuously. In particular, the measured
current-voltage characteristic (CVC) $U\left( I\right) $ is continuous.
However, $U\left( I\right) $ is not smooth at $I=I_{0}$, where $I_{0}$ is
the value of $I$ where the distribution of luminosity over the electrode
surface changes its symmetry. This transition is caused by a
symmetry-breaking bifurcation that occurs at $I=I_{0}$, with stable states
existing on both sides of the bifurcation point.

The above scenario may be illustrated by the following example. If the
discharge vessel is axially symmetric, then the mathematical problem
describing steady-state current transfer to the electrode admits an axially
symmetric (2D) solution, describing the spotless mode of current transfer,
and a 3D solution, describing a mode with a spot. (More precisely, there is
a family of 3D solutions which differ one from the other by the azimuthal
position of the spot. Other families of 3D solutions, describing modes with
several spots, may exist as well.) It is a usual situation that the 3D
spot-mode solution branches off from the 2D spotless-mode solution; a
symmetry-breaking, or pitchfork, bifurcation. Note that a brief summary of
information from the bifurcation theory relevant to this work can be found
in Appendix of \cite{2009f}; a further discussion can be found, e.g., in
reviews \cite{2014b,Trelles2016}.

Let us designate by $I_{0}$ the value of discharge current at which the
bifurcation occurs and assume for definiteness that the 2D solution is
stable for $I>I_{0}$ and unstable for $I<I_{0}$. It may happen that the 3D
solution branches off into the range $I<I_{0}$, where the 2D solution is
unstable; a supercritical bifurcation. According to the general trends of
the bifurcation theory, the 3D solution is stable at least in the vicinity
of the state $I=I_{0}$ in this case. If such a situation is investigated
experimentally and $I$ in the experiment exceeds $I_{0}$, the discharge will
operate in the 2D spotless mode and the luminosity distribution over the
electrode surface will be axially symmetric. As $I$ is reduced down to
values below $I_{0}$, the luminosity distribution starts deviating from
being axially symmetric and the deviation grows proportionally to $\sqrt{%
I_{0}-I}$: a 3D spot starts being formed.

Let us now consider abrupt transitions. The initial and final states may be
of the same or different symmetries, e.g., transitions from a 2D spotless
state to states with a well developed 3D spot or a well developed 2D ring
spot are both included in the consideration. Let us designate by $I_{0}$ the
value of $I$ at which the transition occurs. Since such transitions are
accompanied by jumps in the discharge parameters, the measured CVC $U\left(
I\right) $ is discontinuous at $I=I_{0}$.

There are two possible reasons for abrupt transitions. One of them is the
loss of stability of the mode that existed before a transition, occurring at 
$I=I_{0}$. If an abrupt transition occurs in a monotonic way, i.e., without
temporal oscillations of the electrode luminosity and discharge parameters,
in particular, discharge voltage, then the increment of the perturbations,
against which the stability is lost and which normally have a symmetry lower
than that of the initially existing mode, is real and vanishes at $I=I_{0}$.
The latter means that two steady-state solutions exist in the vicinity of
the state $I=I_{0}$: a solution describing the initially existing mode and a
solution of a lower symmetry, describing the mode with the perturbations.
Hence, a pitchfork bifurcation occurs at $I=I_{0}$. In order to illustrate
this scenario, let us return to the above example and consider the case
where the 3D spot-mode solution branches off into the range $I>I_{0}$, where
the 2D spotless-mode solution is stable; a subcritical bifurcation. In this
case, the 3D solution is usually unstable in the vicinity of the state $%
I=I_{0}$; e.g., Appendix of \cite{2009f}. If the discharge operates in the
spotless mode in the experiment and $I$ is reduced down to values below $%
I_{0}$, the discharge will abruptly switch to another mode and this
switching will occur in a monotonic way, i.e., without temporal oscillations.

Let us now consider the case where an abrupt transition is accompanied by
temporal oscillations. The increment of the perturbations against which the
stability is lost is imaginary at $I=I_{0}$ in this case. Hence, no
steady-state solution bifurcates from the initially existing mode at the
state $I=I_{0}$; i.e., the transition is unrelated to a bifurcation.

The other possible reason of abrupt transitions is that the mode that
existed before the transition has a turning point at $I=I_{0}$. In other
words, this mode has two distinct branches, which exist in the range $I\leq
I_{0}$ (or $I\geq I_{0}$) and merge at $I=I_{0}$, so the mode does not exist
for $I>I_{0}$ (or, respectively, $I<I_{0}$). One can say that the mode has
reached the limit of its existence region at $I=I_{0}$ and turned back; a
fold, or saddle-node, bifurcation. If the discharge operates on one of the
branches of this mode and the current is increased (or, respectively,
decreased), the discharge will abruptly switch to another mode as the value $%
I=I_{0}$ has been reached. Given that the increment of the relevant
instability vanishes at $I=I_{0}$ \cite{2009f}, one can expect that the
switching occurs in a monotonic way.

In summary, quasi-stationary transitions between states with different
symmetries are related to symmetry-breaking (pitchfork) bifurcations of
steady-state solutions; abrupt transitions are related to bifurcations of
steady-state solutions provided that they occur in a monotonic way, i.e.,
without temporal oscillations, and the relevant bifurcations are pitchfork
or fold.

Discharge vessels are axially symmetric in many experiments. Pitchfork
bifurcations of only two types may occur in such configurations \cite%
{2009f,2014b}. First, it is breaking of axial symmetry, i.e., branching of a
3D mode, where the distribution of luminosity over the electrode surface is
periodic in the azimuthal angle with an arbitrary period ($2\pi $, or $\pi $%
, or $2\pi /3$, or $\pi /2$ \textit{etc}), from a 2D mode, where the
distribution of luminosity is axially symmetric. Second, it is doubling of
period with respect to the azimuthal angle, i.e., branching from a 3D mode
with one of the periods $\pi $, $\pi /2$, $\pi /3$, $\pi /4$ \textit{etc} of
a 3D mode with double this period. It follows, in particular, that
transitions with changes of symmetry of other types cannot occur through
pitchfork bifurcations of steady-state solutions, and are always abrupt.

The above general reasoning is valid for mode changes on electrodes of any
dc discharges. In the next sections, this reasoning will be applied to
particular cases of cathodes of arc and dc glow discharges.

\section{Mode transitions on cathodes of arc discharges}

\label{Cathodes of arc discharges}

In the case of refractory cathodes of high-pressure arc discharges, the
theory based on the concept of multiple solutions has gone through a
detailed experimental validation by means of different methods, such as
spectroscopic measurements, electrostatic probe measurements, electrical and
pyrometric measurements, and calorimetry; see, e.g., \cite%
{Dabringhausen2005,Bergner2011,2013a,Schmidt2013}, review \cite{2008c} and
references therein, and also the recent review \cite{Mentel2018}

The theory of current transfer to cathodes of arc discharges is simpler from
the theoretical point of view than the theory for the case of glow
discharge. The eigenvalue problem governing the stability of steady-state
solutions against small perturbations is self-adjoint (Hermitian) in this
case \cite{2007a}. This means, in particular, that the spectrum of
perturbations is real, a conclusion that was confirmed by numerical
calculations \cite{2007d,2009d}. It follows that all abrupt transitions are
monotonic in time. Indeed, no oscillations of arc voltage and luminosity of
the cathode surface is observed in the experiments on transitions between
diffuse and spot modes on arc cathodes; e.g., \cite%
{Boetticher2004,Boetticher2006,2008d,2013a}. Hence, all abrupt transitions
are related to bifurcations of steady-state solutions. In more general
terms, any transition between different modes, be it quasi-stationary or
abrupt, is related to bifurcations of steady-state solutions in the case of
arc cathodes.

In the simplest case of a rod cathode with a flat tip, a 2D diffuse mode of
current transfer occurs in the experiment at high currents and a 3D mode
with a spot at the edge of the cathode occurs at low currents, as
schematically shown by solid lines in Fig.\ 1. (Note that patterns with
several spots have been observed on cathodes of high-pressure arc discharges
in more complex arrangements, such as magnetically rotating arcs \cite%
{Wang2015}.) The transitions between the two modes are shown by the arrows
in Fig.\ 1; they are abrupt without temporal oscillations and manifest
hysteresis. In agreement with the reasoning of Sec.\ \ref{Transitions
between different spot patterns}, these transitions represent an indication
of the presence of pitchfork or fold bifurcations of steady-state solutions.

The latter conclusion may be compared with theoretical results \cite%
{2007a,2007d}. The theory predicts that the diffuse mode and the 3D mode
with a spot at the edge of the cathode are the only modes that contain
stable sections. The stable and unstable sections of each mode are
schematically shown in Fig.\ 1. In the case of the diffuse mode, these
sections are separated by the state $B$, where a subcritical pitchfork
bifurcation occurs. In the case of the spot mode, the stable and unstable
sections are separated by the turning point $K$, where a fold bifurcation
occurs. Thus, the accurate theory indicates that the transition from the
diffuse mode to the spot mode is related to the pitchfork bifurcation and
the return transition is related to the fold bifurcation, in agreement with
the reasoning of Sec.\ \ref{Transitions between different spot patterns}.

\section{Mode transitions on cathodes of dc glow discharges}

\label{Bifurcations on cathodes of dc glow discharges}

\subsection{State-of-the-art of the theory}

In the case of dc glow discharges, multiple solutions have been shown to
exist even in the most basic models and the solutions computed up to now
describe many features of the patterns observed; e.g., \cite{2014b} and
references therein and \cite{2016a}. In particular, the modelling has shown
that self-organization on cathodes of glow microdischarges can occur not
only in xenon, but also in other plasma-producing gases; a prediction which
has been confirmed by subsequent observations of microdischarges in krypton 
\cite{2014c} and argon \cite{Zhu2017}. On the other hand, the comparison
between the theory and the experiment has been merely qualitative up to now.

The eigenvalue problem governing stability of steady-state solutions against
small perturbations is not self-adjoint for glow cathodes. Therefore, the
spectrum of perturbations need not be real. Indeed, a numerical
investigation of stability of 2D modes \cite{2011b} has given a spectrum
that contains both real and complex eigenvalues (and is considerably more
elaborate than the spectrum in the case of arc cathodes). It follows that
abrupt transitions between different spot patterns may be oscillatory, in
contrast to the case of arc cathodes. Note that this conclusion is
consistent with the experiment: for example, temporal oscillations of the
discharge voltage have been observed in the course of transition from the
Townsend to normal discharge \cite{Melekhin1984,Phelps1993,Kaganovich1994}.)
Such transitions are unrelated to bifurcations of steady-state solutions.

\subsection{Analyzing experimental observations}

A\ wealth of self-organized spot patterns and transitions between different
patterns has been observed on cathodes of dc glow microdischarges \cite%
{Schoenbach2004,Moselhy2004,Korolev2005,Takano2006,Takano2006_ICOPS,Lee2007,Zhu2007,Lee2008,Schoenbach2012,2014c,Zhu2014,2016f}%
. It follows from Sec.\ \ref{Transitions between different spot patterns}
that a detailed experimental investigation of transitions between different
spot patterns, performed with sufficiently small steps in $I$ and a
sufficiently high temporal resolution, is needed to unambiguously identify
transitions that are related to bifurcations of steady-state solutions.
Unfortunately, such investigations seem to be absent. The most detailed data
are published in the work \cite{Zhu2014}, where the discharge current was
adjusted on the microampere scale. The question as to whether the observed
transitions are quasi-stationary or abrupt with or without oscillations was
not studied. However, transitions between states of different symmetries
that seem to be continuous (i.e., quasi-stationary) have been observed;
e.g., transitions between states with a large spot occupying the central
part of the cathode and a ring-like arrangement of four spots \cite[Fig.\ 2]%
{Zhu2014}, or between a ring spot and a ring-like arrangement of five spots 
\cite[Fig.\ 5]{Zhu2014}. Question arises as to if these transitions can
occur through pitchfork bifurcations of steady-state solutions, according to
the first scenario described in Sec.\ \ref{Transitions between different
spot patterns}.

In more general terms, one can try to identify in the observations \cite%
{Schoenbach2004,Moselhy2004,Korolev2005,Takano2006,Takano2006_ICOPS,Lee2007,Zhu2007,Lee2008,Schoenbach2012,2014c,Zhu2014,2016f}
all changes of symmetry that may occur through pitchfork bifurcations of
steady-state solutions. There is a possibility that these transitions can be
realized in a quasi-stationary way, although this is not always the case as
exemplified by the transition depicted by the vertical arrow above state $B$
in Fig.\ 1. On the contrary, transitions that are unrelated to pitchfork
bifurcations of steady-state solutions surely cannot be realized in a
quasi-stationary way, i.e., are always abrupt.

The discharge vessels are axially symmetric in most of the above-cited
experiments. If follows from Sec.\ \ref{Transitions between different spot
patterns} that pitchfork bifurcations of only two types may occur in such
configurations: breaking of axial symmetry and doubling of period with
respect to the azimuthal angle. Hence, one should try to identify
transitions with changes of symmetry of these two types in the available
experimental data \cite%
{Schoenbach2004,Moselhy2004,Korolev2005,Takano2006,Takano2006_ICOPS,Lee2007,Zhu2007,Lee2008,Schoenbach2012,2014c,Zhu2014,2016f}%
. If such transitions exist, one should try to find the relevant
bifurcations by means of numerical modelling. If the bifurcations have been
found, one will be able to compare the computed patterns in the vicinity of
the bifurcation points with the patterns observed in the experiments.

Most of the transitions reported in \cite%
{Schoenbach2004,Moselhy2004,Korolev2005,Takano2006,Takano2006_ICOPS,Lee2007,Zhu2007,Lee2008,Schoenbach2012,2014c,Zhu2014,2016f}
do not belong to either of the two above types. None of these transitions
can occur through bifurcations of steady-state solutions, hence these
transitions cannot be realized in a quasi-stationary way. In particular,
this applies to successive transitions between ring arrangements of 4, 5, 6,
5, 4, and 3 spots shown in, e.g., \cite[Fig.\ 2]{Zhu2014} and \cite[Fig.\ 2]%
{Takano2006}. It is interesting to point out that this conclusion is
consistent with the experimental observation that the transition between the
ring arrangements of 6, 5, and 4 spots was irreversible: it could be
realized when the current is lowered, but attempts to increase the current,
when the discharge was operating in these modes, led to the extinction of
the discharge \cite{Takano2006}.

However, transitions that do belong to one of the two possible types of
pitchfork bifurcation (with either a breaking of axial symmetry, or a
doubling of period with respect to the azimuthal angle) have been observed
and are listed in Table I. Note that the two aforementioned transitions
observed in \cite{Zhu2014} that appear to be quasi-stationary (those between
states with a large spot occupying the central part of the cathode and a
ring-like arrangement of four spots \cite[Fig.\ 2]{Zhu2014} and between a
ring spot and a ring-like arrangement of five spots \cite[Fig.\ 5]{Zhu2014})
exhibit a breaking of axial symmetry and therefore can indeed occur through
pitchfork bifurcations; accordingly, these transitions are listed in the
table.

The bifurcation that can be responsible for the second transition in Table I
was encountered in \cite{2016a}. The period-doubling bifurcation that can be
responsible for the fifth transition has been encountered as well, although
for plasma-producing gases different from xenon, which was used in the
experiments \cite{Zhu2014b}: helium \cite[Fig.\ 9]{2013g} and krypton \cite[%
Fig.\ 2]{2014c}. In this work, these bifurcations are numerically
investigated in detail and the computed patterns in the vicinity of the
bifurcation points are compared with the experiment. Also reported in this
work is the finding and analysis of the bifurcation that corresponds to the
third transition in Table I, for which experimental images taken with a very
fine step over discharge current are available \cite[Fig.\ 5]{Zhu2014}.

\subsection{Numerical modelling}

\label{second}

\subsubsection{The models}

\label{The models}

Two numerical models of glow discharges are used in this work, one of them
being basic and the other one more detailed. Both models follow standard
lines. For completeness, a summary of differential equations, boundary
conditions, and data used for transport and kinetic coefficients is given in
Appendix \ref{model}. In brief, the models may be described as follows.

The detailed model comprises equations of conservation of electrons, singly
charged atomic ions, singly charged molecular ions, excimers, and an
effective species for excited atoms that combines all of the excited states
of the $6s$ manifold ($6s[3/2]_{2}$, $6s[3/2]_{1}$, $6s^{\prime }[1/2]_{0}$,
and $6s^{\prime }[1/2]_{1}$), Poisson's equation, and an equation for the
conservation of electron energy. Transport equations for charged-particle
species and electron energy density are written in the drift-diffusion
approximation, transport equations\ for the excited neutral species describe
diffusion. The geometry considered is that of the so-called cathode boundary
layer discharge device, which was used in the vast majority of the
experiments \cite%
{Schoenbach2004,Moselhy2004,Korolev2005,Takano2006,Takano2006_ICOPS,Lee2007,Zhu2007,Lee2008,Schoenbach2012,2014c,Zhu2014,2016f}
and comprises a flat cathode and a perforated anode, separated by a
dielectric, with the radius of the opening in the anode equal to the radius
of the discharge cavity in the dielectric; e.g., Fig.\ 1 of \cite{Zhu2014}.
It is assumed that the charged and excited particles coming from the plasma
are absorbed, and subsequently neutralized and deexcited, respectively, at
the surfaces of the electrodes and the dielectric.

The above-described detailed model is computationally costly and therefore
not suitable for serial 3D simulations, required for the purposes of this
work. It was shown previously experimentally \cite{Takano2006_ICOPS} and
computationally \cite{2016a} that self-organized patterns in the cathode
boundary layer discharge and a discharge with parallel-plane electrode
configuration are qualitatively similar. An account of detailed chemical
kinetics does not produce a qualitative effect as well \cite{2013g}.
Therefore, most of simulations reported below have been performed by means
of a more basic model, which relies on a simple chemical kinetic scheme and
assumes a parallel-plane electrode configuration. (We note right now that
the results obtained in the framework of the basic and detailed models are
qualitatively similar, in agreement with the above.) The basic model takes
into account only one ion species (molecular ions), the only ionization
channel (direct ionization from the ground state by electron impact) with a
rapid conversion of the produced atomic ions into molecular ions, and
employs the local-field approximation (i.e., the electron kinetic and
transport coefficients are treated as known functions of the local reduced
electric field). The discharge vessel is assumed to be a cylinder with the
end faces being the electrodes and the lateral surface being insulating. The
neutralization of the ions and the electrons at the dielectric is neglected,
so particles coming from the plasma are reflected back. Note that the effect
of the neutralization has been well understood by now (e.g., \cite{2016a}
and references therein); as far as 3D spots are concerned, it results in the
migration of spots away from the wall in the direction to the center of the
cathode \cite{2013g}. With this in mind, the assumption of negligible
neutralization is sufficient for most purposes of this work, while making
computations less costly and easier to analyze.

\subsubsection{Identifying the relevant bifurcations}

Results of simulations reported in this section have been obtained by means
of the basic model for the following conditions: a discharge in xenon under
the pressure of $30\,\mathrm{Torr}$, the electron temperature $T_{e}=1\unit{%
eV}$, the heavy-particle temperature $T_{h}=300\unit{K}$\textit{, }the
interelectrode gap and discharge radius both of $0.5\unit{mm}$, and the
secondary electron emission coefficient $\gamma =0.03$.

In order to show the place of the bifurcations being investigated (those
corresponding to the second, third, and fifth transitions in Table I) in the
general pattern of self-organization in dc glow microdischarges, we will
briefly introduce the latter, referring to \cite{2014b} for details. In the
framework of the basic model, the problem admits a 1D solution describing a
mode in which all the variables depend only on the axial variable. This mode
exists at all values of the discharge current and may be termed the
fundamental mode. There are also multidimensional modes which bifurcate
from, and rejoin, the fundamental mode; the so-called second-generation
modes. Fig.\ 2 depicts the current-voltage characteristics (CVC) of the
fundamental mode and the first five second-generation modes. $\left\langle
j\right\rangle $ in this figure is the average current density evaluated
over a cross section of the discharge vessel (which is proportional to the
discharge current). The schematics illustrate distributions of current
density on the cathode surface associated with each mode. $a_{i}$ and $b_{i}$
designate bifurcation points where second-generation modes branch off from
and rejoin the fundamental mode. The modes are ordered by decreasing
separation of the bifurcation points: the mode designated $a_{1}b_{1}$ and
is the one with the bifurcation points positioned further apart, the mode $%
a_{2}b_{2}$ is the one with the second largest separation between
bifurcation points, and so on.

The modes $a_{1}b_{1}$ and $a_{3}b_{3}$ have been computed\ previously (\cite%
{2011a} and \cite{2010a}, respectively) and are included in Fig.\ 1 for the
sake of completeness; we only note that $a_{1}b_{1}$ is 3D with the
azimuthal period of $2\pi $ while $a_{3}b_{3}$ is 2D with one branch
associated with a spot at the center of the cathode and the other branch
with a ring spot at the periphery of the cathode. The other modes, $%
a_{2}b_{2}$, $a_{4}b_{4}$, and $a_{5}b_{5}$, are 3D with periods $\pi $, $%
2\pi /3$, and $\pi /2$, respectively. The evolution with discharge current
of the cathodic spot patterns associated with these modes is shown in Fig.\
3. Let us consider first the evolution of the patterns associated with the
mode $a_{2}b_{2}$; Fig.\ 2(a). The state $151.05\unit{V}$ is positioned in
the vicinity of the bifurcation point $a_{2}$ and the spot pattern comprises
two very diffuse cold spots at the periphery of the cathode. Further away
from $a_{2}$, the cold spots expand and at state $151.79\unit{V}$ start
merging. This corresponds to the retrograde section of the CVC $a_{2}b_{2}$
seen in Fig.\ 2(b) in a narrow current range around $280\unit{A}\unit{m}%
^{-2} $. As current is further reduced towards $b_{2}$, the two cold spots
expand further and the resulting pattern comprises two well-pronounced hot
spots at the periphery; state $160.4\unit{V}$. Note that this pattern is
similar to those observed in the experiment; cf.\ \cite[Fig.\ 5]{Takano2006}%
. Finally, the state $173.93\unit{V}$ is positioned in the vicinity of the
bifurcation point $b_{2}$ and the hot spots are very diffuse.

The patterns associated with the mode $a_{4}b_{4}$ are shown in Fig.\ 3(b).
The state $151.01\unit{V}$ is positioned in the vicinity of the bifurcation
point $a_{4}$ and the pattern comprises three very diffuse cold spots at the
periphery. Further away from $a_{4}$, the spots become better pronounced and
a cold spot appears at the center; states $151.15\unit{V}$ and $151.53\unit{V%
}$. Note that similar patterns with three hot spots have been observed in
the experiment, cf.\ \cite[Fig.\ 5]{Takano2006} and \cite[Fig.\ 2]{Zhu2014}.
As current is further reduced towards $b_{4}$, the cold spot at the center
is gradually transformed into a hot spot. The hot spots become well
pronounced and a pattern comprising three (hot) spots at the periphery and a
central spot is formed; state $151.74\unit{V}$. (It is this pattern which is
shown in the schematic in Fig.\ 1.) Note that patterns with three spots at
the periphery and a spot at the center similar to that of the state $151.74%
\unit{V}$ have also been observed in the experiment, cf.\ \cite[Fig.\ 4]%
{Lee2008}. Note also that the transition between patterns with well-defined
cold and hot spots is not accompanied by retrograde behavior, in contrast
with the case of the mode $a_{2}b_{2}$. The state $172.48\unit{V}$ is
positioned in the vicinity of the bifurcation point $b_{4}$ and the hot
spots are very diffuse.

The evolution of patterns associated with the mode $a_{5}b_{5}$ shown in
Fig.\ 3(c) follows the same trend as the mode $a_{4}b_{4}$. Note that
patterns with four spots at the periphery have also been observed in the
experiment, cf.\ \cite[Fig.\ 2]{Takano2006}, \cite[Fig.\ 2]{Zhu2014}.

A convenient graphical representation, or bifurcation diagram, of the modes $%
a_{4}b_{4}$ and $a_{5}b_{5}$ is given in Fig.\ 4 with the use of the
coordinates $\left( \left\langle j\right\rangle ,j_{c}\right) $, where $%
j_{c} $ is the current density at the center of the cathode. This
representation allows a quick identification of the state where the
switching between patterns comprising cold and hot spots at the center
happens: it is the point in Fig.\ 4 where the line representing the mode in
question intersects the straight line representing the 1D mode. For currents
higher than the one corresponding to the switching, the current density at
the center is lower than that corresponding to the 1D mode and the pattern
comprises a cold spot at the center;\ $j_{c}>\left\langle j\right\rangle $\
for lower currents and the pattern comprises a hot spot at the center.

Breaking of axial symmetry occurring at the state $a_{5}$ (Figs.\ 2 and 4)
corresponds to the second transition in Table I. In order to identify
bifurcations corresponding to the third and fifth transitions, one needs to
consider third-generation modes, i.e., 3D modes that branch off from and
rejoin second-generation modes.

Three third-generation modes bifurcating from the mode $a_{3}b_{3}$,
designated $a_{3,1}b_{3,1}$, $a_{3,2}b_{3,2}$, and $a_{3,3}b_{3,3}$, are
shown in Fig.\ 5. They branch off from and rejoin that branch of the mode $%
a_{3}b_{3}$ which is associated with a ring spot at the periphery; the
bifurcations are breaking of axial symmetry. The modes $a_{3,1}b_{3,1}$, $%
a_{3,2}b_{3,2}$, and $a_{3,3}b_{3,3}$ have the periods of $2\pi /3$, $2\pi
/5 $, and $\pi /3$, respectively, and are associated with spot patterns
comprising three spots at the periphery of the cathode, five spots, and six
spots, respectively. Since none of the patterns shown in Fig.\ 5 comprise a
spot at the center, the coordinates $\left( \left\langle j\right\rangle
,j_{c}\right) $ would be inconvenient and the coordinates $\left(
\left\langle j\right\rangle ,j_{e}\right) $ are used, where $j_{e}$ is the
current density at a fixed point on the periphery of the cathode which
coincides with the center of one of the spots.

The evolution of the spot patterns associated with the mode $a_{3,2}b_{3,2}$
is shown in Fig.\ 6. At state $151.82\unit{V}$, which is positioned near the
bifurcation point $a_{3,2}$, the ring spot is slightly non-uniform in the
azimuthal direction. Further away from $a_{3,2}$, the non-uniformity evolves
into well-pronounced spots (states $151.81\unit{V}$ and $151.84\unit{V}$).
The spots become smaller as the current is further reduced (state $152.26%
\unit{V}$). As the bifurcation point $b_{3,2}$ is approached, the spots
expand once again (state\ $167.94\unit{V}$). In the close vicinity of $%
b_{3,2}$ (state $170.70\unit{V}$) a ring spot with a weak non-uniformity in
the azimuthal direction is seen. Note that patterns similar to those shown
in Fig.\ 6 have been observed in the experiment \cite[Fig.\ 5 ]{Zhu2014}.

The behavior of the modes $a_{3,1}b_{3,1}$ and $a_{3,3}b_{3,3}$ follows the
same trend as the behavior of the mode $a_{3,2}b_{3,2}$. The patterns are
similar to experimentally observed patterns comprising three and six spots
inside the cathode; cf.\ \cite[Figs.\ 2 and 5]{Takano2006} and \cite[Fig.\ 2]%
{Zhu2014}. Note, however, that the pattern with three spots associated with
the mode $a_{3,1}b_{3,1}$ is similar to the pattern with three spots
appearing in some states belonging to $a_{4}b_{4}$ (states $151.15\unit{V}$
and $151.53\unit{V}$ in Fig.\ 3(b)) and it is difficult to know which one of
these two modes was observed in the experiments \cite{Takano2006,Zhu2014}. A
similar comment applies to the mode $a_{3,3}b_{3,3}$.

Breaking of axial symmetry occurring at the states $a_{3,2}$ and $b_{3,2}$
in Fig.\ 5 corresponds to the third transition in Table I. The bifurcation
corresponding to the fifth transition is period doubling occurring at the
state $a_{10,1}$ in Fig. 7. Here $a_{10}b_{10}$ is a second-generation mode
with the period of $\pi /3$ and $a_{10,1}b_{10,1}$ is a third-generation
mode with the period of $2\pi /3$. The period doubling at $a_{10,1}$ occurs
as follows: every second spot gradually moves from the periphery towards the
center of the cathode; eventually a central spot is formed. (Note that the
image illustrating the mode $a_{10,1}b_{10,1}$ in Fig.\ 7 corresponds to the
situation where the central spot has already been formed.) This is similar
to how the similar bifurcation occurs in helium \cite[Fig.\ 9]{2013g} and
krypton \cite[Fig.\ 2]{2014c} except that in krypton the central spot is
already present at the bifurcation point $a_{10,1}$.

\subsection{Comparing the modelling and the experiment}

\label{bif}

As discussed in the preceding section, the bifurcation corresponding to the
second transition in Table I is breaking of axial symmetry occurring at the
state $a_{5}$ (Figs.\ 2 and 4), where a mode with a ring-like arrangement of
four spots (mode $a_{5}b_{5}$) branches off from the (fundamental) mode with
an axially symmetric spot occupying the whole cathode surface except for the
periphery (the abnormal discharge). The bifurcation corresponding to the
third transition is breaking of axial symmetry occurring at the states $%
a_{3,2}$ and $b_{3,2}$ (Fig.\ 5), where a mode with a ring-like arrangement
of five spots (mode $a_{3,2}b_{3,2}$) branches off from the axially
symmetric mode with a ring spot ($a_{3}b_{3}$). The bifurcation
corresponding to the fifth transition is period doubling occurring at the
state $a_{10,1}$ (Fig.\ 7), where the mode $a_{10,1}b_{10,1}$, which is
associated with three spots at the periphery and three spots closer to the
center and has the period of $2\pi /3$, branches off from the mode $%
a_{10}b_{10}$, which is associated with a ring-like arrangement of 6
identical spots at the periphery and has the period of $\pi /3$.

The computed patterns in the vicinity of the bifurcation points are compared
with the patterns observed in the experiments in Fig.\ 8. The experimental
images shown in Figs.\ 8(a) and 8(b) have been taken from Figs.\ 2 and 5,
respectively, of \cite{Zhu2014}. Those shown in Fig.\ 8(c) have been kindly
provided by W. Zhu and P. Niraula \cite{Zhu2014b}; we note for completeness
that the geometry in this experiment was the same as in \cite{Zhu2014}, the $%
\mathrm{Xe}$ pressure was $100\unit{torr}$, and the current and voltage were
nearly the same for both frames: $0.155\unit{mA}$ and $278\unit{V}$.

The first one of the computed images shown in Fig.\ 8(a) represents the
bifurcation point $a_{5}$. The other images correspond to states belonging
to the mode $a_{5}b_{5}$ in the vicinity of $a_{5}$. The last one of the
computed images shown in Fig.\ 8(b) represents the bifurcation point $%
a_{3,2} $, the other images correspond to states belonging to the mode $%
a_{3,2}b_{3,2}$ in the vicinity of $a_{3,2}$. The first one of the computed
images shown in Fig.\ 8(c) represents the bifurcation point $a_{10,1}$, the
other images correspond to states belonging to the mode $a_{10,1}b_{10,1}$
in the vicinity of $a_{10,1}$.

It is seen from Fig.\ 8 that the computed patterns in the vicinity of the
bifurcation points closely resemble the patterns observed in the
experiments. This supports the hypothesis that the transitions between
patterns of different symmetries observed in the experiment and listed in
the second, third, and fifth lines of Table I are quasi-stationary and occur
through pitchfork bifurcations.

The transition between the abnormal discharge and a mode with four spots,
shown in Fig.\ 8(a), resembles the well-known transition between the
abnormal and normal glow discharges, the difference being that there are
four spots in the 3D mode and not just one as in the normal discharge. A
question arises as to what is the reason of this difference and why just
four spots are formed and not two or three. This question is related to a
more general question as to why patterns with multiple spots have been
observed in glow microdischarges but not in regular-scale glows and is of
significant interest.

It is seen from Fig.\ 2 that the mode with four spots at the periphery
branches off from the abnormal discharge, at the state $a_{5}$, through a
supercritical bifurcation, while the modes with one, two, and three spots
branch off, at the states $a_{1}$, $a_{2}$, and $a_{4}$, through subcritical
bifurcations. (Note that this is a typical situation: low- and high-order
second-generation modes tend to branch off through, respectively,
subcritical and supercritical bifurcations \cite[Fig.\ 3]{2014b}.) As
discussed in Sec.\ \ref{Transitions between different spot patterns}, a
usual necessary condition for a quasi-stationary transition between two
steady-state modes connected by a pitchfork bifurcation is that the
bifurcation be supercritical. Therefore, it may seem that the modelling
results shown in Fig.\ 2 explain why the abnormal discharge in the
experiment with microdischarges goes into the mode with four (rather than
one, two, or three) spots, as seen in Fig.\ 8(a). On the other hand, the
experimental CVC of this transition \cite[Fig.\ 3a]{Zhu2014} apparently
represents a diagram of a subcritical bifurcation, and so does also the CVC\
shown in \cite[Fig.\ 3a]{Takano2006}. Thus, there is a discrepancy between
the measurements, on one hand, and numerical modelling and the usual trend
of the bifurcation theory, on the other. The other discrepancy between the
measured and computed CVCs is that the discharge voltage in the 3D mode is
lower than that in the (axially symmetric) abnormal mode in the experiment
but slightly higher in the modelling.

In order to try to resolve the discrepancies, the bifurcations occurring at
the states $a_{4}$ and $a_{5}$ have been recomputed by means of the detailed
model, described in Sec.\ \ref{The models} and Appendix \ref{model}; Fig.\
9. Since the geometry of the discharge vessel and the boundary conditions
describing absorption of the charged particles at the wall invalidate the 1D
solution, the role of the fundamental mode (abnormal discharge) is played by
the first 2D mode \cite{2014b}. All second-generation solutions are 3D, and
the first four modes, which have the periods of $2\pi $, $\pi $, $2\pi /3$,
and $\pi /2$, respectively, are designated $a_{1}b_{1}$, $a_{2}b_{2}$, $%
a_{4}b_{4}$, and $a_{5}b_{5}$ (i.e., the designation $a_{3}b_{3}$ is skipped
in order to maintain consistency with the designations of the
second-generation modes computed in the framework of the basic model).

It is seen from the schematics in Fig.\ 9 that the account of neutralization
of the charged particles at the wall of the vessel causes the spots to
migrate from the edge to the inside of the cathode. The discharge voltage
computed in the framework of the detailed model is significantly higher than
the voltage in the basic model. On the other hand, the patterns of the CVCs
are similar, in agreement with what was said in Sec.\ \ref{The models}. In
particular, the bifurcation occurring at $a_{5}$ is supercritical while the
one at $a_{4}$ is subcritical (as well as those at $a_{1}$ and $a_{2}$,
which are not shown in Fig.\ 9) and the discharge voltage in the 3D modes in
both models is slightly higher than that in the fundamental mode. Thus, the
above-described discrepancies have not been resolved and further
computational and experimental work is needed.

\section{Summary and the work ahead}

\label{Conclusion}

The existence, or not, of bifurcations of steady-state solutions describing
different modes of current transfer to electrodes of dc discharges is
related to the underlying physics and is therefore of significant interest.
Bifurcations manifest themselves in the experiment as transitions between
modes with different spot patterns, which occur as the discharge current $I$
is varied. Two scenarios of such transitions are possible: (1)
quasi-stationary (continuous) and, consequently, reversible transitions
between states where distributions of luminosity over the electrode surface
possess different symmetries and (2) transitions that occur abruptly even
for very small variations of $I$. Quasi-stationary transitions are related
to a symmetry-breaking (pitchfork) bifurcation. If the discharge vessel is
axially symmetric, pitchfork bifurcations of only two types may occur:
breaking of axial symmetry and doubling of period with respect to the
azimuthal angle. Abrupt transitions that occur in a monotonic way, i.e.,
without temporal oscillations, are related to pitchfork or fold
bifurcations. Finally, abrupt transitions accompanied by temporal
oscillations are unrelated to a bifurcation of steady-state solutions.

The above general reasoning is valid for mode changes on electrodes of any
dc discharges. In the case of (refractory) cathodes of high-pressure arc
discharges, the eigenvalue problem governing stability of steady-state
solutions against small perturbations is self-adjoint and its spectrum is
real. Therefore, all abrupt transitions are monotonic in time, in agreement
with what is known from the experiment. It follows that any transition
between different modes, be it quasi-stationary or abrupt, is related to a
bifurcation of steady-state solutions in the case of arc cathodes. Thus,
transitions between diffuse and spot modes of current transfer, frequently
observed in the experiment, represent an indication of the presence of
pitchfork or fold bifurcations of steady-state solutions, as predicted by
the theory.

A\ wealth of spot patterns and transitions between different patterns have
been observed on cathodes of dc glow microdischarges \cite%
{Schoenbach2004,Moselhy2004,Korolev2005,Takano2006,Takano2006_ICOPS,Lee2007,Zhu2007,Lee2008,Schoenbach2012,2014c,Zhu2014,2016f}%
. In particular, transitions between states of different symmetries that
seem to be continuous (i.e., quasi-stationary) have been observed in \cite%
{Zhu2014}. It is legitimate to hypothesize that such transitions occur
through pitchfork bifurcations (breaking of axial symmetry or period
doubling) of steady-state solutions according to the first above-mentioned
scenario. This hypothesis has been confirmed by numerical modelling: the
relevant bifurcations have been found and the computed patterns in the
vicinity of the bifurcation points are found to closely resemble the
patterns observed in the course of the transitions in the experiment. Note
that new 3D modes of current transfer were computed in the course of finding
the bifurcations and these new modes are associated with experimental spot
patterns reported in the literature.

Thus, available experimental data on multiple modes of current transfer to
cathodes of dc glow and arc discharges provide clear indications of the
presence of pitchfork or fold bifurcations of steady-state solutions, as
predicted by the theory. While the comparison between the theory and the
experiment still remains qualitative in the case of dc glow cathodes, the
agreement is convincing and lends further support to the theory.

A detailed experimental investigation of transitions between different spot
patterns on cathodes of glow microdischarges, performed with sufficiently
small steps in $I$ and a sufficiently high temporal resolution and
accompanied by numerical modelling, would allow verification of the above
scenarios. For example, it would be very interesting to verify the
theoretical prediction that successive transitions between ring arrangements
of 4, 5, 6, 5, 4, and 3 spots shown in, e.g., \cite[Fig.\ 2]{Zhu2014} and 
\cite[Fig.\ 2]{Takano2006} cannot be realized in a quasi-stationary way, in
contrast to the second, third, and fifth transitions in Table I. It should
be stressed that this prediction is consistent with the experimental
observation that the transition between the ring arrangements of 6, 5, and 4
spots was irreversible \cite{Takano2006}. Note also that the theory predicts
that any transition, except those between an axially symmetric mode and a 3D
mode and those between 3D modes with doubling of azimuthal period, will be
abrupt even for very small variations of discharge current and/or voltage.

Another interesting question to be addressed is the one discussed in the end
of Sec. \ref{bif} and concerns the discrepancy between the measured CVC of
the transition from the abnormal discharge and the mode with four spots, on
the one hand, and numerical modelling as well as the usual trend of the
bifurcation theory, on the other.

\textbf{Acknowledgments} $\;$The work was supported by FCT - Funda\c{c}\~{a}%
o para a Ci\^{e}ncia e a Tecnologia of Portugal through the project
Pest-OE/UID/FIS/50010/2013. The authors are grateful to Dr.\ WeiDong Zhu and
Mr.\ Prajwal Niraula for discussion of the experiment \cite{Zhu2014} and
kindly providing the experimental images shown in Fig.\ 8(c).\newpage

\appendix{}

\section{Equations and boundary conditions}

\label{model}

The system of differential equations describing the detailed model reads%
\begin{eqnarray}
\nabla \cdot \mathbf{J}_{e} &=&S_{e},\;\;\;\mathbf{J}_{e}=-D_{e}\,\nabla
n_{e}+n_{e}\,\mu _{e}\,\nabla \varphi ;  \notag \\
\nabla \cdot \mathbf{J}_{\varepsilon } &=&e\mathbf{J}_{e}\cdot \nabla
\varphi -S_{\varepsilon },\;\;\;\mathbf{J}_{\varepsilon }=-D_{\varepsilon
}\,\nabla n_{\varepsilon }+n_{\varepsilon }\,\mu _{\varepsilon }\,\nabla
\varphi ;  \notag \\
\nabla \cdot \mathbf{J}_{i\beta } &=&S_{i\beta },\;\mathbf{J}_{i\beta
}=-D_{i\beta }\,\nabla n_{i\beta }-n_{i\beta }\,\mu _{i\beta }\,\nabla
\varphi ;  \notag \\
\nabla \cdot \mathbf{J}_{ex\beta } &=&S_{ex\beta },\;\ \ \mathbf{J}_{ex\beta
}=-D_{ex\beta }\,\nabla n_{ex\beta };  \notag \\
\varepsilon _{0}\,\nabla ^{2}\varphi  &=&-e\,(n_{i1}+n_{i2}-n_{e}).
\label{1}
\end{eqnarray}%
Here $\beta =1,2$; the indexes $e,\varepsilon ,i1,i2,ex1,ex2$ refer to
electrons, electron energy density, atomic ions, molecular ions, atoms in
excited states, and excimers, respectively; $n_{\alpha }$, $\mathbf{J}%
_{\alpha }$, $D_{\alpha }$, $\mu _{\alpha }$, $S_{\alpha }$ are the number
density, density of the transport flux, diffusion coefficient, mobility, and
rate of production of particles per unit time and unit volume of species $%
\alpha $ ($\alpha =e,i1,i2,ex1,ex2$); the electron energy density is defined
as $n_{\varepsilon }=n_{e}\overline{\varepsilon }$, where\ $\overline{%
\varepsilon }$ is the average electron energy, and coincides with $3/2$ of
the electron pressure; $\mathbf{J}_{\varepsilon }$ is the density of
electron energy flux; $D_{\varepsilon }$ and $\mu _{\varepsilon }$ are the
so-called electron energy diffusion coefficient and mobility; $%
S_{\varepsilon }$ is the rate of loss of electron energy per unit time and
unit volume due to collisions; $\varphi $ is electric potential; $%
\varepsilon _{0}$ is permittivity of free space, and $e$ is the elementary
charge. The transport coefficients used are the same as in \cite{2013g}. The
set of reactions comprises electron impact ionization from the ground state,
an effective excited atomic state, and the metastable molecular state;
excitation of atoms by electron impact;\ atomic ion conversion to molecular
ions with neutral atoms playing the role of the third body; metastable
pooling;\ dissociative recombination;\ photon emission from the atomic and
molecular excited states;\ conversion of excited atoms to excimers with
neutral atoms playing the role of the third body. The kinetic coefficients
are the same as in \cite{2013g}.

The computation domain corresponds to the cathode boundary layer discharge
device (e.g., Fig.\ 1 of \cite{Zhu2014}). Let us introduce cylindrical
coordinates $\left( r,\phi ,z\right) $ with the origin at the center of the
cathode and the $z$-axis coinciding with the axis of the vessel. Then the
computation domain is a union of the cylinder $\left\{
r<R,\;0<z<h_{d}+h_{a}\right\} $ and the half-space $\left\{
z>h_{d}+h_{a}\right\} $, where $h_{d}$ and $h_{a}$ are thicknesses of the
dielectric and the anode and $R$ is the radius of the opening in the anode
and of the cavity in the dielectric. The circle $\left\{ r<R,\;z=0\right\} $
is the surface of the cathode, $\left\{ r=R,\;h_{d}<z<h_{d}+h_{a}\right\}
\cup \left\{ r>R,\;z=h_{d}+h_{a}\right\} $ is the surface of the anode, and
the surface $\left\{ r=R,\;0<z<h_{d}\right\} $ is dielectric.

The boundary conditions for Eqs.\ (\ref{1}) are written in the conventional
form (e.g., \cite{Hagelaar2000, Salabas2002}):%
\begin{eqnarray}
J_{\alpha z} &=&\frac{1}{4}\sqrt{\frac{8k_{B}T_{\alpha }}{\pi m_{\alpha }}}%
n_{\alpha }\text{$,$ }J_{ez}\mathbf{=}-\gamma J_{iz}-\gamma J_{i2z}\mathbf{,}%
\ J_{\varepsilon z}\mathbf{=(}E_{I}-2W\mathbf{)}J_{ez},\text{ }\varphi =0%
\text{\thinspace };  \notag \\
J_{\alpha n} &=&\frac{1}{4}\sqrt{\frac{8k_{B}T_{\alpha }}{\pi m_{\alpha }}}%
n_{\alpha }\text{$,\ $}J_{en}=\frac{1}{2}\sqrt{\frac{8k_{e}T_{e}}{\pi m_{e}}}%
n_{e}\text{$,$ }J_{\varepsilon n}=\frac{1}{2}\sqrt{\frac{8k_{B}T_{e}}{\pi
m_{e}}}n_{\varepsilon }\text{$,$ }\varphi =U;  \notag \\
J_{\alpha r} &=&\frac{1}{4}\sqrt{\frac{8k_{B}T_{\alpha }}{\pi m_{\alpha }}}%
n_{\alpha }\text{$,$}\;J_{er}=\frac{1}{2}\sqrt{\frac{8k_{e}T_{e}}{\pi m_{e}}}%
n_{e},\text{$\ $}J_{\varepsilon r}=\frac{1}{2}\sqrt{\frac{8k_{B}T_{e}}{\pi
m_{e}}}n_{\varepsilon }\text{$,$ }J_{i1r}+J_{i2r}-J_{er}=0,  \label{2}
\end{eqnarray}%
at the surface of the cathode, anode, and dielectric, respectively. Here $%
\gamma $ is the effective secondary emission coefficient, which is assumed
to characterize all mechanisms of electron emission (due to ion, photon, and
excited species bombardment) \cite{Raizer1991}; $E_{I}$ is the ionization
energy of the incident ions;\textit{\ }$W$ is the work function of the
cathode surface; $U$ is the discharge voltage; the subscripts $r$, $z$, and $%
n$ denote radial, axial, and normal projections of corresponding vectors
(the normal $n$\textbf{\ }points outwards from the computation domain); $%
\alpha =i1,i2,ex1,ex2$; $k_{B}$ is Boltzmann's constant; $T_{\alpha }=T_{h}$%
, where $T_{h}$ is the heavy-particle temperature (a given parameter); $%
T_{e}=2\overline{\varepsilon }/3k_{B}$ is the electron temperature; and $%
m_{\alpha }$ are the particle masses.

Results of simulations performed by means of this model, reported in this
work, refer to the following conditions: discharge in xenon under the
pressure of $75\,\mathrm{Torr}$, $T_{h}=300\unit{K}$, $h_{d}=0.25\unit{mm}$, 
$h_{a}=0.25\unit{mm}$, $R=0.375\unit{mm}$, and $\gamma =0.075$.

The system of differential equations describing the basic model comprises
the first and last equations in Eq.\ (\ref{1}) and the third equation
written for only one ion species. Boundary conditions at the cathode, anode,
and dielectric are written as, respectively,%
\begin{eqnarray}
z &=&0:\;\;\;\frac{\partial \,n_{i}}{\partial \,z}=0,\text{\quad }%
J_{ez}=-\gamma J_{iz}\text{$,$\quad }\varphi =0\text{\thinspace };  \notag \\
z &=&h:\;\;\;n_{i}=0\text{$,$\quad }\frac{\partial \,n_{e}}{\partial \,z}=0%
\text{\thinspace $,$\quad }\varphi =U;  \notag \\
r &=&R:\;\;\;\frac{\partial \,n_{i}}{\partial r}=\frac{\partial \,n_{e}}{%
\partial r}=0\,,\text{\quad }J_{ir}-J_{er}=0,  \label{3}
\end{eqnarray}%
where $n_{i}$ and $\mathbf{J}_{i}$ are the number density and transport flux
density of the ions. The first boundary condition in the first line and the
second boundary condition in the second line imply that diffusion fluxes of
the attracted particles at the electrode surfaces are neglected compared to
drift. The first and second boundary conditions in the third line imply that
the neutralization of the ions and the electrons at the dielectric is
neglected. The transport and kinetic coefficients used in the basic model
for xenon are taken from \cite{2010a}.

Numerical results reported in this work have been computed using the
commercial finite element software COMSOL Multiphysics. The detailed model
was implemented using the Plasma module of COMSOL, which was adapted so that
it could be used in combination with a stationary solver and supplemented
with a mesh element size based stabilization method to reduce the P\'{e}clet
number. The basic model was implemented using the Transport of Diluted
Species module, and also solved using a stationary solver. \textit{\newpage }

\bibliographystyle{apsrev4-1}
\bibliography{Matt_thesis}
\vspace{0.5in}

\begin{center}
\begin{tabular}{|c|c|c|c|c|}
\hline
\multicolumn{2}{|c|}{Higher-symmetry mode} & \multicolumn{2}{|c|}{
Lower-symmetry mode} & Source \\ \hline
Symmetry & Pattern & Symmetry & Pattern &  \\ \hline
2D & Central spot & 3D, $\pi $ & 2 symmetric spots & \cite[Fig.\ 5]%
{Takano2006} \\ \hline
2D & Central spot & 3D, $\pi /2$ & Ring of 4 spots & \cite[Fig.\ 2]%
{Takano2006}, \cite[Fig.\ 2]{Zhu2014} \\ \hline
2D & Ring spot & 3D, $2\pi /5$ & Ring of 5 spots & \cite[Fig.\ 5]{Zhu2014}
\\ \hline
2D & 
\begin{tabular}{l}
Central spot, \\ 
ring spot%
\end{tabular}
& 3D, $2\pi /5$ & 
\begin{tabular}{l}
Central spot, \\ 
ring of 5 spots%
\end{tabular}
& \cite[Fig.\ 6]{Zhu2014} \\ \hline
3D, $\pi /3$ & 
\begin{tabular}{l}
Ring of \\ 
6 spots%
\end{tabular}
& 3D, $2\pi /3$ & 
\begin{tabular}{l}
2 rings of 3 \\ 
spots each%
\end{tabular}
& \cite{Zhu2014b} \\ \hline
\end{tabular}%
\smallskip

Table I. Transitions between modes with different spot patterns observed on
cathodes

of glow microdischarges that are potentially related to bifurcations. Numbers

in the columns 'Symmetry' in cases of 3D modes designate azimuthal period.
\end{center}

\end{document}